\documentclass[pre,aps,showpacs,nofootinbib,twocolumn]{revtex4}
\usepackage{graphicx}

\begin{document}

\title{Nucleosome repositioning via loop formation}
\author{I.M. Kuli\'{c} and H. Schiessel}
\affiliation{%
Max-Planck-Institut f\"{u}r Polymerforschung, Theory Group, POBox
3148, D 55021 Mainz, Germany
}%

\date{\today}

\title{Nucleosome repositioning via loop formation}

\begin{abstract}
Active (catalysed) and passive (intrinsic) nucleosome
repositioning is known to be a crucial event during the
transcriptional activation of certain eucaryotic genes. Here we
consider theoretically the intrinsic mechanism and study in detail
the energetics and dynamics of DNA-loop-mediated nucleosome
repositioning, as previously proposed by Schiessel et al. (H.
Schiessel, J. Widom, R. F. Bruinsma, and W. M. Gelbart. 2001. {\it
Phys. Rev. Lett.} 86:4414-4417). The surprising outcome of the
present study is the inherent nonlocality of nucleosome motion
within this model -- being a direct physical consequence of the
loop mechanism. On long enough DNA templates the longer jumps
dominate over the previously predicted local motion, a fact that
contrasts simple diffusive mechanisms considered before. The
possible experimental outcome resulting from the considered
mechanism is predicted, discussed and compared to existing
experimental findings.
\end{abstract}


\maketitle

\section{Introduction}

\begin{figure}
\includegraphics*[width=8cm]{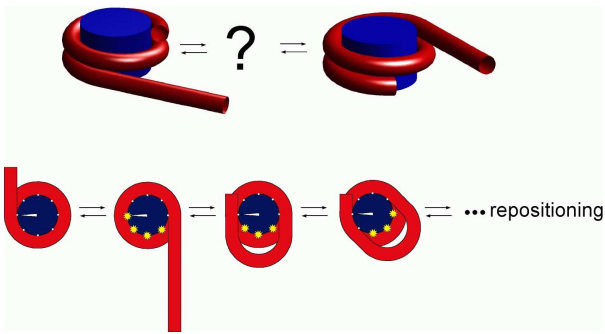}
\caption{The basic problem setting: how does the histone-octamer
move along the DNA template? Below: the DNA loop mechanism as
proposed in in Ref.~\cite{Schiessel}}
\end{figure}

The nucleosome, the most abundant DNA-protein complex in nature,
is the basic unit of eucaryotic chromatin organization. It is
roughly a cylinder of 6nm height and 10nm diameter, consisting of
a protein octamer core and 147 basepairs (bp) of DNA tightly
wrapped around it in 1 and 3/4 left-handed superhelical turns. The
genes of all higher organisms, ranging from simple ones like yeast
to most elaborate like humans are all organized in long arrays of
nucleosomes with short DNA segments (linkers) of 50-100 bp
interpolating between them, comparable to a beads-on-a-string
chain \cite{Widom Review,Kornberg Review,Wolffe Book}. The higher
order organization of these units, being most probably a solenoid-
or zig-zag, crossed-linker- like fiber with 30 nm diameter is
still under great dispute though it received increasing
theoretical and experimental support in recent years. Above that
scale of organization, the higher order structures which link the
30 nm to the final ''big X'' like structure, the packed
chromosome, are still unknown. Though there are several
biologically motivated speculations about the ''big X'' its
definite structure remained a long lasting puzzle for the last 20
years, defying all biophysical, biochemical and molecular genetics
efforts to resolve it because of its intrinsic softness and
fuzziness.

An additional obstacle for understanding the chromatin structure
is the fact that it is highly dynamic on all organization scales.
Starting at the macroscopic chromosome level we see that its
structure can strongly vary throughout the cell cycle on
timescales of hours or days. Below that on timescales of seconds
and minutes, the structure of the 30 nm fiber itself is subjected
to great variations due to transcription, replication, biochemical
modification and other dynamic processes. Finally, at the lowest
organization level, the nucleosome itself has been shown to be a
dynamical structure being moved along the DNA by chromatin
remodeling complexes on expense of ATP \cite{Remodeling Review
1,Remodeling Review2}. Interestingly, it was experimentally
observed \cite{Beard,Spadafora,Meersseman,Pennings} that
nucleosomes can move even autonomously on short DNA\ segments.
This intrinsic repositioning behavior was shown to be strongly
temperature dependent. At room temperature it occurs roughly on
timescales of $\sim $1 hour indicating the existence of
significant energetic barriers. Besides the fact that the
repositioning does indeed occur and is of intramolecular nature
(the nucleosome stays on the same DNA segment) the underlying
scenario could not be figured out. It was speculated by Pennings
et. al. \cite{Meersseman,Pennings} that the mechanism was some
type of nucleosome-sliding or screwing motion. An alternative
explanation which appears to be more consistent with the discrete
jumps and large barriers observed by Pennings et. al. has been
recently proposed in Ref.~\cite{Schiessel}. In this model the
basic step in the repositioning process is a partial unwrapping of
DNA\ from the very ends of the nucleosome \cite{Polach
Widom,Anderson Widom} followed by a backfolding of DNA with a
small 10 bp mismatch (cf. Fig.~1). The result of this process is
the formation of a small DNA bulge or loop on the octamer surface.
Once trapped on the nucleosome surface this small defect carrying
some discrete quantum of DNA extra length (a multiple of 10 bp,
the DNA repeat length) can propagate by diffusion in both
directions. If the loop happens to surround the nucleosome and
comes out at the opposite side (in respect to where it was
created) the nucleosome is eventually repositioned by a distance
given by the ''pulled in'' extra length. The energetic barrier and
rates of repositioning were computed and were shown to be
consistent with the Pennings et al. experiment
\cite{Meersseman,Pennings}. Moreover, the 10 bp discrete step
repositioning observed in the experiment (discrete bands, no 1 bp
spaced intermediates) came out as a natural consequence of the
loop length quantization. The latter is enforced by the strongly
preferred DNA minor groove - octamer interaction and the discrete
binding sites at the nucleosome surface as deduced from the
crystallographic structures \cite {Luger}.

In Ref.~\cite{Schiessel} small loops with short excess length of
typically $\sim 1-2\times $10 bp were considered and it was shown
that the looping energies involved increase rapidly with the
excess length implying that only the shortest (10 bp) loop
contributes significantly to the repositioning mechanism.
Consequently the model predicts a classical discrete random walk
with a jump-size of 10 bp -- instead of a 1 bp motion that would
be implied by sliding/cork-screwing mechanism. Apart from the
discrepancy in the elementary step size, both models predict very
similar behavior: a local one-dimensional diffusive motion along
the DNA chain.

In this paper we will carefully reanalyze the idea of
loop-mediated repositioning by applying the classical tool of the
Kirchhoff kinetic analogy which provides us with analytic
solutions of the loop problem and enables\ us to look at loops of
virtually any given excess length. The main outcome of our study
will be a different picture of repositioning which physically
results from the looping mechanism: on short up to moderately long
segments of up to 2-3$\times $ $l_{P}$ ($l_{P}$: DNA persistence
length) the repositioning is a jumpy process with largest possible
loops being the most dominant ones in contrast to short 10 bp
steps as conjectured before. For longer\ and very long (infinite)
DNA segments there is an optimal jump size of order $\sim
O(l_{p})$ and the behavior is superdiffusive in contrast to the
previously predicted diffusive mechanism. As we will see below,
these predictions allow us to clearly distinguish between
different repositioning mechanisms in experiments expected to be
performed in near future \cite{Stephanie}.

\section{Energetics of Loops}

\begin{figure}
\includegraphics*[width=8cm]{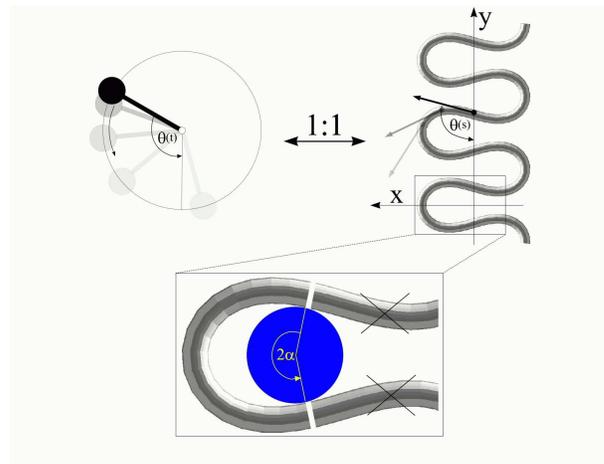}
\caption{The Kirchhoff kinetic analogy between the spinning top
and the bent/twisted rod depicted for a special case: the plane
pendulum - planar rod equivalence. The inset shows how an
intranucleosomal loop can be constructed by inscribing the octamer
(gray disk) into the bent rod. The nucleosome opening angle
$2\alpha $ accounts for the adsorption energy cost (see text for
details).}
\end{figure}

Let us now consider the energetics of an intranucleosomal DNA
loop. We will describe it within the framework of the
Euler-Kirchhoff theory for the static equilibrium of rods (Fig.
2). For simplicity and because of the approximate planarity of the
problem we can in first approximation assume the nucleosome and
the loop-forming DNA to be in one plane and the DNA to be free of
any twisting deformation. It this case the free energy of our
system is simply divided into two components, the planar elastic
DNA-bending and a histone-octamer DNA\ interaction:
\begin{equation}
U_{tot}=U_{bend}+U_{ads}  \label{Etot}
\end{equation}
The bending energy (within the linear elasticity approximation)
can be written in terms of the local DNA curvature $\kappa $
\begin{equation}
U_{bend}=\frac{A}{2}\int_{-L/2}^{L/2}\kappa ^{2}\left( s\right) ds
\label{Ebend}
\end{equation}
with $A \approx 50$ $nm\cdot k_{B}T$ being the bending rigidity of
DNA at room temperature and physiological salt concentrations
\cite{Hagerman}. The rod is assumed to be parametrized by its
contour length parameter $s$ ranging from $-L/2$ to $L/2$ with
$L$\ being the total length of the loop. The latter can be
expressed in terms of two independent quantities: the excess
length $\Delta L$ and the nucleosome opening angle $\alpha $
(Fig.2)
\begin{equation}
L\left( \alpha ,\Delta L\right) =2\alpha R+\Delta L  \label{L}
\end{equation}
where $R\approx 4$ nm is the effective nucleosome radius, or more
precisely the distance from the center of the nucleosome to the
central DNA axis. Because the DNA\ can enter the nucleosome only
in quantized orientations (with its minor groove phosphates) and
bind only to discrete positions on the protein surface
\cite{Luger}, the excess length $\Delta L=n\times h_{DNA}$ is to a
good approximation an integer multiple of the DNA repeat length
$h_{DNA} = 3.4nm$.

The second part in the total energy Eq.~\ref{Etot} $U_{ads}$ comes
from the (predominantly electrostatic) interaction between the
positively charged protein surface and the negatively charged DNA.
It can be roughly measured from experiments probing the
competitive protein binding to nucleosomal DNA \cite{Polach
Widom,Anderson Widom}. Neglecting the discreteness of charges
(binding sites) on the histone octamer surface it can in first
approximation be assumed to be proportional to the opening angle
$\alpha $
and the adsorption energy density $\varepsilon _{ads}$%
\begin{equation}
U_{ads}=2\alpha R\varepsilon _{ads}  \label{Einter}
\end{equation}
with $\varepsilon _{ads}\approx 0.5-1.0$ $k_{B}T/nm$ as roughly
extracted
from \cite{Polach Widom} \footnote{%
In Eq.~\ref{Einter} we assume that the interaction is only short
ranged (contact interaction) which is justified by the very short
Debye screening length of \ $\approx 1nm$ under physiological salt
conditions.}. Here and in the
following we assume an intermediate value of $\varepsilon _{ads}=0.7$ $%
k_{B}T/nm$.

\subsection{Ground states of trapped loops}

In order to compute the ground state for a trapped
intranucleosomal loop we have to consider shapes that minimize the
total energy \ref{Etot} under two constraints:

\begin{enumerate}
\item  The excess length $\Delta L$ is prescribed. Therefore we
have the
relation Eq.~\ref{L} between the opening angle and the total loop length $L$%
\begin{equation}
\Delta L=L-2\alpha R=const.  \label{Constraint1}
\end{equation}

\item  At the two ends $s_{\pm }=\pm L/2$
the rod has to be tangential on an inscribed circle of given
radius
(representing the nucleosome) \footnote{%
Because of the symmetry we have to impose the conditions only on
one side.}:
\begin{equation}
R=\left| \frac{y\left( \frac{L}{2}\right) }{-x^{\prime }\left( \frac{L}{2}%
\right) }\right| =const.  \label{Constraint2}
\end{equation}
\end{enumerate}

Here $x\left( s\right) $ and $y\left( s\right) $ are the Cartesian
coordinates of the rod axis as a function of the arc-length
parameter $s$ (cf. Fig.~2). The absolute value in the second
constraint needs to be introduced formally for dealing with
crossed rod solutions (which we consider later on) and can be
omitted for simple uncrossed loops.

For an analytical description it is convenient to use the angle $\theta $ $%
=\theta \left( s\right) $ between the DNA tangent and the $y$ axis
(cf. Fig.~2) as a variable describing the DNA\ centerline. In this
case the integrated sine (cosine) of $\theta $ over the arc-length
parameter $s$ gives the x (y) Cartesian coordinate of any point
along the rod, and the squared derivative $\left( \theta ^{\prime
}\right) ^{2}$ gives the rod curvature $\kappa $. Furthermore the
nucleosome opening angle $\alpha $ is simply related to $\theta $
at the boundary
\[
\alpha =
{\theta \left( L/2\right) \text{ for simple loops}
\atopwithdelims\{. \pi -\theta \left( L/2\right) \text{ for crossed loops}}%
\]
The two constraints Eq.~\ref{Constraint1} and Eq.
\ref{Constraint2} can be rewritten in terms of $\theta $ and then
be introduced into the minimization by two Lagrange multipliers
$\mu _{1/2}$. We then arrive at the following lengthy functional
\begin{eqnarray}
\widehat{U}_{tot} &=&A\int_{0}^{L/2}\left( \theta ^{\prime }\right) ^{2}ds%
\text{\ }+2\alpha R\varepsilon _{ads}  \nonumber \\
&&+\mu _{1}\left[ L-(\Delta L+2\alpha R)\right]  \nonumber \\
&&+\mu _{2}\left[ \int_{0}^{L/2}\cos \theta ds-R\sin \alpha
\right] \label{EnergyFunctionalFull}
\end{eqnarray}
Here the first line is the bending + adsorption energy
contribution, the second and third line are the imposed length and
tangency constraint. Eq.\ref {EnergyFunctionalFull} can be
rearranged in a more familiar form
\begin{equation}
\int_{0}^{L/2}\left( A\left( \theta ^{\prime }\right) ^{2}+\mu
_{2}\cos \theta \right) ds+\text{b.t.}
\label{EnergyFunctionalSimplified}
\end{equation}
Here b.t. denotes the boundary terms (depending on $\theta \left(
L/2\right) $ only) that obviously do not contribute to the first
variation inside the relevant $s$ interval. The integral in
Eq.\ref{EnergyFunctionalSimplified} is
evidently analogous to the action integral of the plane pendulum with $%
A\left( \theta ^{\prime }\right) ^{2}$ corresponding to the
kinetic and $-\mu _{2}\cos \theta $ to the potential energy of the
pendulum. The latter analogy is a rather unspectacular observation
knowing the celebrated Kirchhoff's kinetic mapping between
deformed rods and the spinning top, which contains our present
problem as a simple special case. The Kirchhoff's analogy states
that the {\it equilibrium conformations} of (weakly) deformed thin
rods can be mapped to the {\it time-dynamics} of a heavy symmetric
spinning top subjected to a gravitational force. It has been
repeatedly applied (with or without direct reference to Kirchhoff)
to DNA related problems during the last 20 years (e.g. see
\cite{Benham1,Benham2,Le Bret 1,Le Bret 2,Swigon,Shi,Fain Rudnick
1,Fain Rudnick 2,Schiessel2}). For a nice visual review on the
spinning top-elastic rod analogy the reader is referred to Ref.
\cite{Goriely} where the general solutions together with a
''kinetic dictionary'' (time $t\longleftrightarrow $ length
parameter $s$, gravitational force$\longleftrightarrow $rod
tension $\mu _{2}$, axis of revolution$\longleftrightarrow
$tangent vector etc.) are also provided.

The nice thing about Kirchhoff's analogy apart from its esthetical
content is that it provides us with explicit expressions for DNA\
shapes subjected to twist, bending and various geometric /
topological constraints. In our simple planar and twistless case,
the ''spinning top'' simply reduces to the simple plane pendulum.
The corresponding planar and twistless rods, also called the {\it
Euler elastica}, are most generally given by
\begin{equation}
\cos \theta \left( s\right) =1-2m%
\mathrm{sn}^{2}\left( \frac{s}{\lambda }\mid m\right)
\label{GenSol}
\end{equation}
which can be integrated to obtain the general planar rod shape in
Cartesian coordinates:

\begin{mathletters}
\begin{eqnarray}
x\left( s\right) &=&2\sqrt{m}\lambda \mathrm{cn}\left(
\frac{s}{\lambda }\mid m\right)  \label{CartesianX} \\ y\left(
s\right) &=&2\lambda E\left( \frac{s}{\lambda }\mid m\right) -s
\label{CartesianY}
\end{eqnarray}
\end{mathletters}
with sn, dn, cn$(.\mid m)$ being the Jacobi elliptic functions
with the parameter $m$ and

\begin{equation}
E\left( u\mid m\right) :=\int_{0}^{u}\mathrm{dn}^{2}\left( v\mid
m\right) dv
\end{equation}
denoting the incomplete elliptic integral of the second kind in
its
''practical'' form\footnote{%
Some useful formulas and relations for the eliptic functions and
integrals are briefly sketched in \cite{Goriely} and found in
\cite{Abramowitz Stegun} in full depth.}. The two parameters $m>0$
and $\lambda >0$ in Eqs. \ref {CartesianX}, \ref{CartesianY}
characterize the shape and the scale of the solution,
respectively. These solutions are up to trivial plane rotations,
translations, reflections and shifting of the contour parameter $%
s\rightarrow s+s_{0}$ the most general planar Euler elastica
corresponding to the plane pendulum. For different parameters $m$
one obtains different rod shapes corresponding to different
solutions of the spinning top (plane pendulum) motion
\cite{Goriely}. The case $m=0$ describes a pendulum at rest
corresponding to a straight rod, for $0<m<1$ one has strictly
oscillating pendulums corresponding to point symmetric rod shapes which for $%
m<0.\,\allowbreak 92$ are free of self intersections like the one
depicted in Fig.~2. For $m$ higher than $0.\,\allowbreak 92$ the
rods show varying complexity with a multitude of
self-intersections and for $m=1$ one has the so-called homoclinic
pendulum orbit corresponding to a rod solution having only one
self intersection and becoming asymptotically straight for
$s\rightarrow \pm \infty $ (for details see Ref.~\cite{Goriely}).
For even higher values \footnote{%
Usually the parameter $m$ is artificially assumed to be confined
to $0\leq m\leq 1$ but by the Jacobi's real transform for elliptic
functions \cite{Abramowitz Stegun} they stay well-defined even for
$m>1$.} of $m$, i.e., for $m\geq 1$ we have revolving pendulum
orbits corresponding to rods with self-intersections lacking point
symmetry. Finally, the limiting case $m\rightarrow \infty $
corresponds to the circular rod shape.

In order to describe a trapped loop we need to use Eqs.
\ref{CartesianX} and \ref{CartesianY} imposing the constraints Eq.
\ref{Constraint1} and Eq.~\ref {Constraint2}. It turns out to be
more convenient to replace the parameter
set $(\lambda ,m,L)$ with the new (but equivalent) set $(\lambda ,m,\sigma :=%
\frac{L}{2\lambda })$ where we introduced the new dimensionless
parameter $\sigma $
which we call the ''contact parameter''\footnote{%
A more visual parameter set $\left( \alpha ,m,\lambda \right) $
using the opening angle $\alpha =\alpha (\sigma ,m)$ produces
technical problems with non-uniqueness of loop representation.}.
From Eq.~\ref{Constraint2} together with \ref{CartesianX} and
\ref{CartesianY} we can immediately extract the scaling parameter
$\lambda $ and the opening angle in terms of the contact
parameter $\sigma $ and the shape parameter $m$%
\begin{eqnarray}
\lambda \left( \sigma ,m\right) &=&R\left| \frac{%
\mathrm{sn}\left( \sigma \mid m\right) \mathrm{dn}\left( \sigma
\mid m\right) }{2E\left( \sigma \mid m\right) -\sigma }\right|
\label{Lambda} \\
\alpha \left( \sigma ,m\right) &=&\arccos \left[ \pm \left( 2%
\mathrm{dn}^{2}\left( \sigma \mid m\right) -1\right) \right]
\label{Alpha} \\ \pm &:&=sign\left( 2E\left( \sigma \mid m\right)
-\sigma \right) \label{+-}
\end{eqnarray}
Plugging this into Eq.~\ref{Constraint1} we obtain the final form
of the implicit constraint
\begin{eqnarray}
\frac{\Delta L}{2R} &=&\sigma \left| \frac{%
\mathrm{sn}\left( \sigma \mid m\right) \mathrm{dn}\left( \sigma
\mid m\right) }{2E\left( \sigma \mid m\right) -\sigma }\right|
\label{ConstraintFinal} \\
&&-\arccos \left[ \pm \left( 2%
\mathrm{dn}^{2}\left( \sigma \mid m\right) -1\right) \right]
\nonumber
\end{eqnarray}
The curvature $\kappa \left( s\right) $ and the bending energy Eq.
\ref{Ebend} follow from the explicit solution Eq.~\ref{GenSol} to
be
\begin{eqnarray}
\kappa \left( s\right) &=&\frac{2\sqrt{m}}{\lambda }%
\mathrm{cn}\left( \frac{s}{\lambda }\mid m\right)
\label{Curvature} \\
U_{bend} &=&\frac{4mA}{\lambda }\int_{0}^{\sigma }%
\mathrm{cn}^{2}\left( t\mid m\right) dt  \label{EbendExplicit} \\
&=&\frac{4A}{\lambda }\left[ \left( m-1\right) \sigma +E\left(
\sigma \mid m\right) \right]
\end{eqnarray}
The latter expression together with Eqs. \ref{Etot}, \ref{Einter}
- \label{+-} gives a lengthy expression for the total energy with
the sign chosen $\pm $ as in Eq.~\ref{+-}.
\begin{eqnarray}
U_{tot}\left( \sigma ,m\right) &=&  \nonumber \\
&&\frac{4A}{R}\left| \frac{\left[ 2E\left( \sigma \mid m\right)
-\sigma \right] \left[ E\left( \sigma \mid m\right) +\left(
m-1\right) \sigma \right]
}{%
\mathrm{sn}\left( \sigma \mid m\right) \mathrm{dn}\left( \sigma
\mid m\right) }\right|  \nonumber \\
&&+2R\varepsilon _{ads}\arccos \left[ \pm \left( 2%
\mathrm{dn}^{2}\left( \sigma \mid m\right) -1\right) \right]
\label{EtotFinal}
\end{eqnarray}
Now our problem of finding the ground state loop for given excess length $%
\Delta L$ reduces to a two variable ($\sigma ,m$) minimization of
Eq.~\ref {EtotFinal} under the constraint Eq.
\ref{ConstraintFinal}. This final step is easily performed
numerically.

\section{Loop Zoology: Simple and Crossed Loops}

\begin{figure}
\includegraphics*[width=8cm]{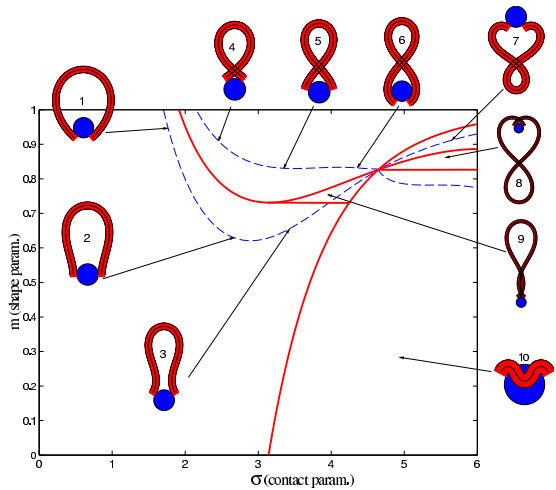}
\caption{The set of possible ground-state solutions is
characterized by two parameters, the contact point parameter
$\sigma $ and the loop shape parameter $m$. Solutions with
constant excess length $\Delta L$ (here $10\times 3.4nm$) are
located along the dashed lines (e.g. loops 1-7). The solid lines
separate loops with different geometric characteristics: simple
(1,2,3), crossed (4,5,6) and ''exotic'' (7,8,9,10) loop shapes.}
\end{figure}

\begin{figure}
\includegraphics*[width=8cm]{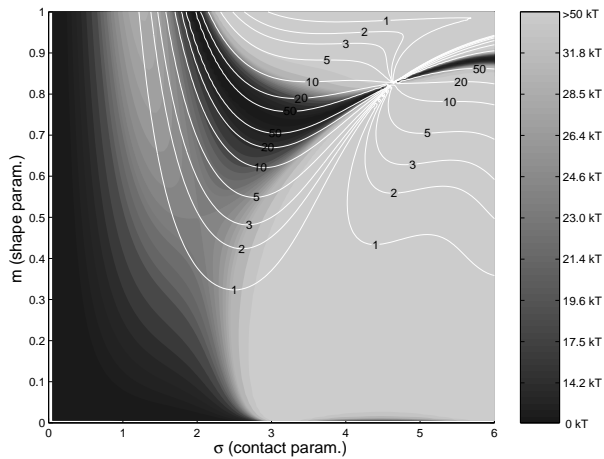}
\caption{Density plot of the total loop energy Eq.~\ref{EtotFinal}
(grayscale levelsets) as a function of $\sigma$ and $m$ (same
parameter range as in Fig. 3). The white contours denote lines of
constant excess length $\Delta L=1,2,3,5,10,20,50\times 3.4$ nm.
For given excess length the ground state is the point on the
corresponding white line with the darkest background (note the
different branches for given $\Delta L$). The parameters are
$\varepsilon _{ads}=0.7 k_{B}T/nm$ and $A=50nm \times k_{B}T$ and
$R=4nm$.}
\end{figure}

We can scan now through the $\sigma -m$ parameter plane and look
at the shapes of the solutions and their energies. In Fig.~3 we
see a small (but most important) part of the whole parameter space
and the corresponding different loop geometries. The dashed lines
indicate parameter values which lead to constant excess length
$\Delta L=10\times 3.4nm$ (corresponding to 100 bps) in accordance
with the constraint Eq.~\ref{ConstraintFinal}. The shapes 1-7 are
examples of 100bp-loops with different geometries. The whole
parameter plane is subdivided by separation lines (solid) into
regions of structurally different solutions. The large region
starting at $\sigma =0$ contains exclusively simple loops (like
1,2 and 3) without self-intersections and nucleosome penetration.
Above that simple-loop-region we find loops with a single
self-intersection (4,5,6) and to the right the loops penetrate the
nucleosome, like loop 10. There are also three other regions with
single and double crossing points (7,8,9) where the loop can also
be on the ''wrong'' side of the nucleosome like in 7 and 8.

We are interested in the energy minimizing loops and the
underlying minimal energies as functions of the excess length
$\Delta L$. A density plot of these energies as function of the
parameters $\sigma $ and $m$ together with the corresponding lines
of constant $\Delta L$ (with $\Delta L=1,2,..,50\times 3.4 nm$) is
given in Fig.~4. As can be seen from Fig.~3 there are, for a given
$\Delta L$, different branches of $(\sigma ,m)$ values
corresponding to uncrossed, simply crossed and other exotic
structures. Of all these structures for short excess lengths,
$\Delta L$ $\lesssim 20\times 3.4nm$, the energetically dominant
ones are simple (uncrossed) loops which we study first. Loops with
larger excess length form crossed structures and are studied in
Section 3.2.

\begin{figure}
\includegraphics*[width=8cm]{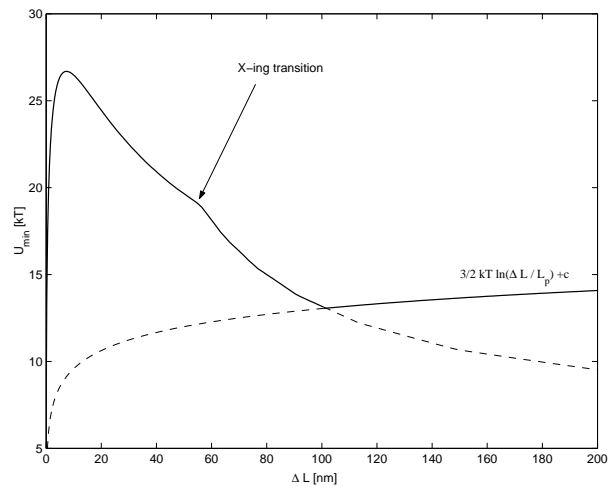}
\caption{The ground state loop energy plotted vs the excess length
$\Delta L$. Note the energy maximum occuring for shorter loops.
For much longer loops (around $\Delta L=60 nm$) a transition from
simple uncrossed to crossed loop shapes occurs leading to a kink
in $U_{\min }\left( \Delta L\right)$. In the regime of low $\Delta
L\lesssim l_{P}$ the elastic energy prevails strongly over entropy
whereas for large loops the entropy starts to dominate the
behavior producing a shallow energy minimum in the cross-over
regime which roughly defines the predominant loop size.}
\end{figure}

\subsection{Simple Loops}

For simple uncrossed loops it is a straightforward numerical task
to minimize Eq.~\ref{EtotFinal} under the constraint of constant
excess length, Eq.~\ref{ConstraintFinal}. For $\varepsilon
_{ads}=0.7k_{B}T/nm$ and all the other parameters as above
($A=50nm \times k_{B}T,$ $R=$ $4nm$) the ground state
energy $U_{min}$ as a function\footnote{%
Formally the quantisation condition $\Delta L=1,2,...\times 3.4nm$
holds as mentioned above. Nevertheless for clarity we consider the
values in between as well.} of the excess length $\Delta L$ is
shown in Fig.~5 (for $\Delta L\lesssim 60nm$; for longer $\Delta
L$-values crossed loops are more favorable as discussed in the
next section). Remarkably we find that the loop energy is
non-monotonous: For small $\Delta L$ $U_{min}$ increases with
$\Delta L$ as $\left( \Delta L\right) ^{1/3}$ (in accordance with
Ref.~\cite{Schiessel} where only small loops were studied). At
some critical excess length $\Delta L =\Delta L_{crit}$ (which is
approximately $\Delta L_{crit}\approx 2.2\times 3.4nm$ for
$\varepsilon _{ads}=0.7$ $k_{B}T/nm$) the loop energy reaches a
maximum (here $U_{min}(\Delta L_{crit})\approx 26k_{B} T$). Beyond
that the energy decreases with increasing $\Delta L$.

In the following we show how this behavior can be explained on the
basis of the loop geometry. Naively one might argue as follows:
For excess lengths shorter than the persistence length of DNA it
is increasingly difficult to store additional length into the loop
because it requires increasing DNA deformation. On the other hand,
for loops longer than $l_P$ the bending energy contribution
becomes very small and hence one expects such ground state loops
relaxing with increasing $\Delta L$. However the reason for
occurence of a maximum of $U_{min}$ around $\approx 2$ excess DNA
lengths, a value which is considerably smaller than the
persistence length, is not obvious. In order to understand this
finding one has to go beyond the simple handwaving heuristics and
needs to take a close look at the details of the loop geometry.

To this end we introduce here a simple approximation technique
which leads to explicit expressions which can be more easily
handled than the exact yet complicated expressions given above. We
call this method the {\it circle-line approximation} and give a
detailed exposition in the Appendix. As we will see this method is
quite accurate and at the same time very intuitive.

\begin{figure}
\includegraphics*[width=8cm]{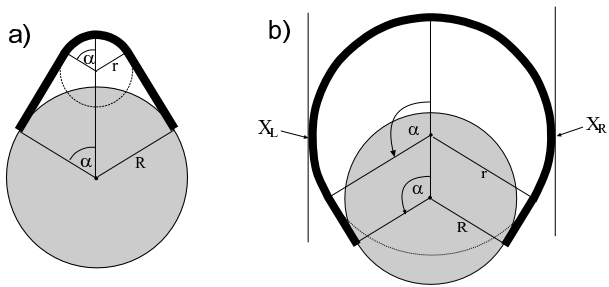}
\caption{Two generic types of simple loop geometries (in the
circle-line approximation): a) the subcritical loop with opening
angle $\alpha $ $<\pi /2$ and b) the supercritical loop with $%
\alpha $ $>\pi /2$. In the former case the introduction of further
excess length leads to an energy increase but in the latter case
to a relaxation of stress: The introduction of additional length
at points $X_{L}$ and $X_{R}$ followed by a relaxation of the
structure obviously decreases the total energy.}
\end{figure}

Looking at the geometrical shapes of the loops in Fig.~3 we notice
that each of them is subdivided into several sections of very high
and very low curvature (cf. also Eq.\ref{Curvature}). In first
approximation we replace the high curvature regions by sections of
circles, the low curvature regions by straight lines (cf. Fig.~6).
Furthermore, to keep the smoothness we assume that the lines are
tangents to the circles. Generally in order to have reasonable
approximations of all possible loop shapes we would need to
consider compositions of several circles and lines (cf. for
instance loops 3, 6, 7). However, if the adsorption energies are
not to high, i.e., if the opening angle $\alpha $ is ''soft
enough'' and does not impose such a severe bending like in loop 3,
such multiply bent loops will not be relevant as ground state
solutions. As it turns out for our problem we already obtain a
quite good approximation by assuming that the loop consists of a
{\it single} circular arc and two lines only. It is characterized
by two quantities : 1) the arc radius $r$ and 2) the nucleosome
opening angle $\alpha $ (cf. Fig.~6 and Appendix). With these
assumptions and after some elementary geometry the constraint Eq.%
\ref{Constraint1} becomes simply
\begin{equation}
\Delta L=2\left( R-r\right) \left( \tan \alpha -\alpha \right)
=const. \label{CostraintCircular}
\end{equation}
Note that the (more complex) second constraint Eq.
\ref{Constraint2} is eliminated through the ''ansatz'' {\it per
se}. The total loop energy is given in terms of the loop radius
$r$ and the opening angle $\alpha $
\[
U_{tot}\left( \alpha ,r\right) =A\frac{\alpha }{r}+2\alpha
R\varepsilon _{ads}
\]
and by applying the constraint Eq.\ref{CostraintCircular} (which
this time can be solved explicitly!) we obtain $U_{tot}$ in terms
of $\alpha \ $and
given $\Delta L$%
\begin{equation}
U_{tot}\left( \alpha \right) =2\alpha \left( A\frac{\tan \alpha -\alpha }{%
2R\left( \tan \alpha -\alpha \right) -\Delta L}+R\varepsilon
_{ads}\right) \label{EtotCircular}
\end{equation}
which is explicit in $\alpha $. We note that this approximation
for $U_{tot}$ is only reasonable for $2R\left( \tan \alpha -\alpha
\right) >\Delta L$, i.e., for not too small $\alpha $ (vs. $\Delta
L$), otherwise the bending contribution diverges or becomes even
negative (the latter is obviously absurd). The reason for this is
that for very small angles $\alpha $
(compared to $\Delta L$) uncrossed\footnote{%
In contrary for crossed loops there still are solutions for small
$\alpha $ (cf. next chapter).} circle-line loops cannot exist for
geometrical reasons. There this most basic approximation breaks
down and we would have to approximate the loop by more than one
circular segment. But as mentioned above, such loops ($\alpha $
small compared to $\Delta L$) are not candidates for the ground
state for moderate $\varepsilon _{ads}\sim O\left( 1\right) $, and
we therefore dispense with giving a discussion of this case.

The nice thing about Eq.~\ref{EtotCircular} is that despite its
simplicity and approximate nature it reproduces the position of
the maximum in Fig.~5 quite well. We find the condition for the
critical excess length $\Delta L_{crit}$ from a simple geometric
distinction between two loop shapes: the subcritical loop (Fig.
6a) with its tangents not being parallel to the $Y$ axis $(\alpha
=0)$ and the supercritical loop (Fig.~6b) having two or more
tangents parallel to the line $\alpha =0$. Suppose now we add
excess length to a subcritical loop by keeping the angle $\alpha
=const$. Obviously the loop-energy increases because the loop
radius r becomes smaller. On the other hand in the supercritical
case we have the opposite situation: the loop energy decreases
with increasing $\Delta L$. This is simply because we could cut
the loop at two points ($X_{L}$ and $X_{R}$ in Fig.~6), introduce
there the additional length (without changing the energy) and then
relax the shape by letting it evolve to the new equilibrium while
keeping $\alpha =const.$ Thus we can obtain the condition for the
critical excess length $\Delta
L_{crit}$ by assuming that the corresponding minimum $\alpha ^{\min }$ of $%
U_{tot}$ just crosses the critical line $\pi /2$ line, i.e.,
$\alpha ^{\min
}\left( \Delta L^{crit}\right) \stackrel{!}{=}\pi /2$ for the searched $%
\Delta L_{crit}$.
\begin{equation}
\left. \frac{d}{d\alpha }\right| _{\alpha =\pi /2}U_{tot}\left(
\alpha \right) \stackrel{!}{=}0
\end{equation}
which can be solved for $\Delta L^{crit}$%
\begin{equation}
\Delta L^{crit}=\frac{4R}{\pi }+\frac{8R^{3}}{\pi A}\varepsilon
_{ads} \label{Lcrit}
\end{equation}
The latter can now be inserted in Eq.~\ref{EtotCircular} leading
to
\begin{equation}
U_{tot}^{crit}=\frac{\pi A}{2R}+\pi R\varepsilon _{ads}
\label{Ucrit}
\end{equation}
For the given values of $R,A,\varepsilon _{ads}$ ($R=4nm$ , $A=50$ $nmk_{B}T$%
, $\varepsilon _{ads}=0.7$ $k_{B}T/nm$)\ we obtain $\Delta
L^{crit}=7.\,\allowbreak 37$nm and $U_{tot}^{crit}=\allowbreak
28.\,\allowbreak 4k_{B}T$ which is in satisfactory agreement with
the exact
numeric results ($\Delta L^{crit}=7.\,\allowbreak 19nm$, $%
U_{tot}^{crit}=\allowbreak 26.\,\allowbreak 7k_{B}T$). More
generally, for not to high adsorption energies ($\varepsilon
_{ads}=0.5-2.0$ $k_{B}T/nm$)\ the circle-line approximation works
well and Eqs. \ref{Lcrit} and \ref{Ucrit} reproduce the exact
positions of the critical point typically with a 5-15\% accuracy.

For an explicit parametric representation of the minimal energy
curve within the circle-line approximation, which in particular
implies the upper results, the reader is referred to the appendix
where the usefulness of this approach is also demonstrated for
some other examples.

\subsection{Crossed and Entropic Loops}

A closer inspection of Fig.~4 shows that the ground state of loops
switches from simple uncrossed loops to crossed loops when one
reaches an excess length around 50 nm. However, as can be seen for
the crossed structures 4, 5 and 6 in Fig.~3 these loops have a
self-penetration at the crossing point. Therefore, a planar theory
is in principle not sufficient to describe such structures. One
possible formal cure for this problem would be to leave the plane
and to consider the rod's self-contacts with the corresponding
point-forces etc. in 3D as done by Coleman et al. in a general
theory of rod self-contacts \cite{Coleman}. However such a
procedure leads to a significant loss of transparence, not only
because of the third dimension entering the scene but also due to
the necessity to subdivide the rod into different regions with
different forces acting in each of them. Instead of following
Coleman at al. \cite{Coleman} we decided to treat the
self-interaction in a perturbative manner as follows. If the
self-contact point is not too close to the nucleosome the rod is
not severely deflected out of the plane by its self-interaction.
Thus it remains roughly planar with some out of plane bending in
$Z$-direction of the rod sections between the nucleosome and the
crossing point. This will cost some additional bending energy
$U_{def}$ that is roughly given by (cf. Appendix)

\begin{equation}
U_{def}\left( \sigma ,m\right) =\frac{2A}{R}\frac{\rho \arctan \left( \frac{%
\rho \tan \alpha \left( \sigma ,m\right) }{\tan ^{2}\alpha \left(
\sigma ,m\right) -\rho ^{2}}\right) }{\tan ^{2}\alpha \left(
\sigma ,m\right) -\rho ^{2}}  \label{Udef}
\end{equation}
Here $\rho :=d/R$ with $d\approx 1nm$ is the DNA radius. We
neglect the slight twisting of the rod induced by the
non-planarity of the DNA and consider the bending only. The
deflection energy Eq.~\ref{Udef} can be phenomenologically
incorporated into the model by simply adding it to
Eq.~\ref{EtotFinal} as a correction term to obtain the final form
of the total energy $U_{tot}^{\ast }$
\[
U_{tot}^{\ast }\left( \sigma ,m\right) =\left\{
\begin{array}{c}
U_{tot}\left( \sigma ,m\right) \text{ for uncrossed (simple)
loops} \\ U_{tot}\left( \sigma ,m\right) +U_{def}\left( \sigma
,m\right) \text{ for crossed loops}
\end{array}
\right.
\]
With this additional modification of $U_{tot}$ we computed
numerically the minimal energy (ground state) solution for any
given excess length $\Delta L$. The graph of the ground state
energy versus $\Delta L$ is shown if Fig.~5. We find that even
with the inclusion of the out-of-plane deflection there is still a
critical length $\Delta L_{cross}$ (here $\approx 60nm$) where the
crossed loops become energetically more favorable than the simple
uncrossed. This behavior that we call the ''crossing transition''
can be rationalized by noting that for long enough loops the
adsorption energy (proportional to $\alpha $) starts to dominate
over the bending energy so that loops with smaller $\alpha $
become increasingly favorable. From the critical length $\Delta
L_{cross}$ on, the gain in adsorption energy (by diminishing
$\alpha $) is more than sufficient to outweigh the (slight)
increase in bending energy together with the additional
self-interaction term, Eq.~\ref{Udef}.

Increasing the length even further we leave the elastic energy
dominated regime in which the entropic effects can be neglected
due to short loop length ($\lesssim $ persistence length). For
larger lengths entropic effects become more and more important and
we ultimatively enter the entropic loop regime. The crossover
between these two regimes is hard to handle analytically
\cite{Yamakawa Book}; for the case of closed loops a perturbative
description has been given in Ref.~\cite{Yamakawa Stockmayer}. For
our purpose it is sufficient only to consider the asymptotic
behavior. In the large loop limit where the loop is longer than
several $l_P$ the chain looses its ''orientational memory''
exponentially and behaves roughly as a random walk which starts
from and returns to the same point. The entropic cost for gluing
the ends of a random walk (long loop) together is then given by
\begin{equation}
U=3/2k_{B}T\ln (\Delta L/l_{P})+E_{0}+S_{0}  \label{Uenropic}
\end{equation}
The first constant, $E_{0}\approx 6.5$ $k_{B}T$ is the bending +
adsorption energy contribution of the overcrossing DNA segments
leaving / entering the nucleosome which can be determined by
minimizing the crossed loop energy (cf. Appendix Eq.~\ref
{A2}) for $\Delta L\rightarrow \infty $. The second additive constant $%
S_{0}\sim O(k_{B}T)$ accounts for the entropic contribution of
DNA-histone octamer interaction volume (the proximity necessary
for the histone octamer and DNA to see each other). Although the
latter constant is not easy to estimate the following prediction
is not sensitive to any additive constant.
We expect a free energy minimum to occur at the overlap between the elastic (%
$\Delta L\lesssim l_{P}$) and entropic ($\Delta L\gg l_{P}$)
region where the decreasing elastic energy is overtaken by the
increasing entropic contribution.

The free energy, Eq.~\ref{Uenropic}, leads to an algebraically
decaying probability $w\left( \Delta L\right)$ for the jump
lengths scaling as $w\sim \left( \Delta L\right) ^{-3/2}$. In
general, power law distributions of the form $w\sim \left( \Delta
L\right) ^{-\gamma }$ with $\gamma>1$ lead to superdiffusive
behavior of the random walker (here the nucleosome). According to
Levy's limit theorem the probability distribution of the random
walker (more precisely, the distribution of the sums of
independent random variable drawn out from the same probability
distribution $w\sim \left( \Delta L\right) ^{-\gamma }$) converges
to a stable Levy distribution of index $\gamma-1$
\cite{Bouchaud,Klafter,Sokolov}. This so-called Levy-flight
differs in many respects from the usual diffusion process as for
short time intervals big jumps are still available with
significant probability. Moreover, all moments (besides possibly
the first few ones) diverge. For our case $\gamma =3/2$ even the
first moment does not exist. We note that the value $3/2$ is based
on the assumption of an ideal chain (no excluded volume); in
general the excluded volume leads to self-avoiding-walk statistics
with a slighly larger value of $\gamma$ around $2.2$
\cite{Sokolov} (cf. also Ref.~\cite{Cloizeaux}). In that case one
has a finite value of the first moment, i.e., of the average jump
length.

\section{The Dynamics of Nucleosome Repositioning}

In the preceding sections we have computed the typical energies
involved in the formation of arbitrary sized loops. Assuming that
a slow creation followed by a fast termal migration of loops
around the nucleosome is the governing mechanism for nucleosome
repositioning we start now considering the repositioning dynamics.
In order to describe the time-dependent evolution of the
nucleosome position we consider its probability distribution along
a DNA\ segment of a length $N\times 10bp$ and write the master
equation governing the jump process
\begin{equation}
\frac{d}{dt}p_{i}=\sum_{j=1,j\neq
i}^{N}w_{ji}p_{j}-p_{i}\sum_{j=1,j\neq i}^{N}w_{ij}
\label{MasterEQ}
\end{equation}
where $p_{i}$ is the probability for the nucleosome being at the admissible%
\footnote{%
Spaced by a multiple of 10 bp from the initial position} position
$i$ on the DNA segment. The transition rate matrix
$\underline{\underline{W}}=(w_{ij})$ is given by
\begin{equation}
w_{ij}=\left\{
\begin{array}{c}
C_{A}\exp \left( -\frac{1}{k_{B}T}U_{\min }\left( h_{D}\left|
i-j\right|
\right) \right) \text{ for }i\neq j \\
-\sum_{k=1,k\neq i}^{N}w_{ij}\text{ for }i=j
\end{array}
\right.   \label{TransitionProbab}
\end{equation}
where $h_{D}=3.4nm$ (DNA helical pitch). $C_{A}$ denotes the
Arrhenius constant involved in the loop formation process that has
in principle to be determined experimentally. The rough estimate
of $C_A^{-1}=10^{-6}s$ is provided in Ref.~\cite{Schiessel} where
it was shown that $C_A$ is essentially given by the inverse
lifetime of the loop (denoted by $A$ in that paper). This means
that typical repositioning times range from seconds to hours.

The (formal) explicit solution of Eqs. \ref{MasterEQ},
\ref{TransitionProbab} together with the previously obtained
minimal energy $U_{\min }$ is given by
\[
\underline{p}\left( t\right) =\exp (\underline{\underline{W}}t)\underline{p}%
\left( 0\right)
\]
The latter solution can now be considered for different cases: for
short or long DNA chains and for the nucleosome placed in the
middle or at the end of the chain.
\begin{figure}
\includegraphics*[width=8cm]{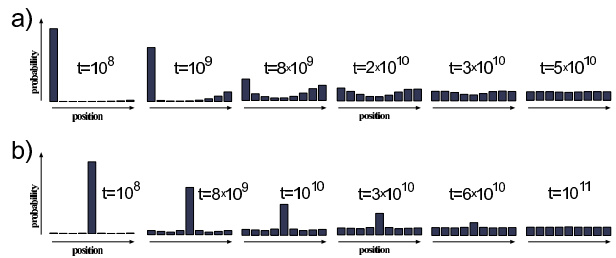}
\caption{Relaxation dynamics of two initial states of nucleosome
positions\ on a short DNA segment (147 + 90 bp): a) the\
nucleosome starting from an end and b) the nucleosome starting
from the middle position. The time unit is the inverse Arrhenius
activation factor $C_{A}^{-1}$ (compare text).}
\end{figure}

\begin{figure}
\includegraphics*[width=8cm]{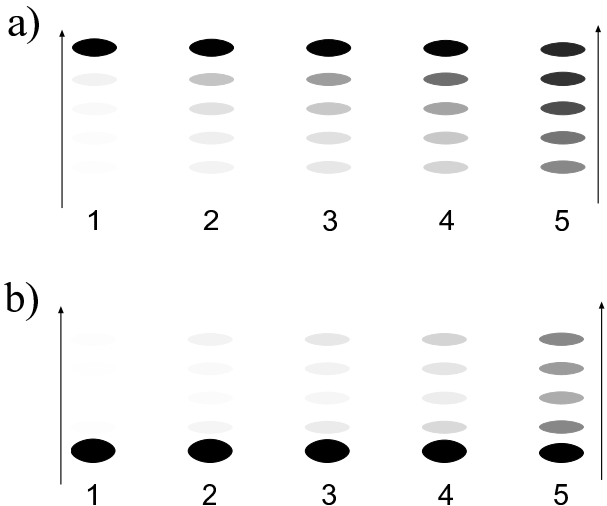}
\caption{Typical (1-D) gel electrophoresis signatures expected for
the relaxation dynamics of the two species from Fig.~7: a)\
nucleosome starts from an end and b) from the middle position. The
lanes 1-5 correspond to incubation times (1,5,10,20,100)$\times
10^{8}$ $C_{A}^{-1}$ respectively. Note: the population of distant
bands in b) lanes 2-4 occurs first, in sharp contrast to what we
expect from a simple (local) diffusive behavior.}
\end{figure}

For short DNA segments we expect a slow repositioning rate due to
high energies involved in small loop formation. In Fig.~7 we
depict the repositioning of a nucleosome on a DNA piece of a
length 147+90 bp. Starting from an end positioned nucleosome (Fig.
7a) we observe a behavior that is
completely unlike a local diffusion mechanism:\ the jumps bigger than $%
\approx 2\times 3.4nm$ start to dominate over the smaller local
ones, which follows from the loop formation energy cf. Fig.~5.
Consequently, in the initial phase of repositioning (of such an
end-positioned population)\ the nucleosomes will predominantly
jump between the two end positions. Later, on a much larger
timescale they gradually start to explore the positions towards
the middle of the DNA segment. Could we extract such a behavior
from an experiment using gel-electrophoretic separation (as in
\cite{Meersseman}, \cite{Pennings})? The basis of such separations
is the fact that the gel-electrophoretic mobility of nucleosomes
on DNA pieces (longer than 147bp) increases roughly linearly with
its distance from the middle position, i.e., DNA pieces with the
nucleosome sitting close to the end run much faster in gels than
equivalent middle positioned nucleosomes do. We can exploit this
(empirical) fact to mimic the outcome of a gel-electrophoresis
experiment (cf. Figs. 8 and 10). In Fig.~8a we depict such a
simulated gel pattern for the middle positioned nucleosome. Since
symmetric species are not distinguished by this experimental
method and are projected onto the same bands (symmetric left/right
positions lead to the same mobility), the expected nonlocality of
motion cannot be extracted from the structure of the bands.

For the same short segment, but with the nucleosome starting from
the middle position (Fig.~7b) the situation is slightly
different:\ the neighboring positions are\ populated more
homogeneously, although there is a small initial underpopulation
of the $2\times 3.4nm$ distant position as expected from the
energy maximum occurring there. In this case, a slight initial
''population gap'' can be observed in gel electrophoresis (Fig.
8b) which in this case would be sufficient to distinguish between
a jumpy and a diffusive behavior, since the latter would obviously
lack the ''population gap''.

\begin{figure*}
\includegraphics*[width=16cm]{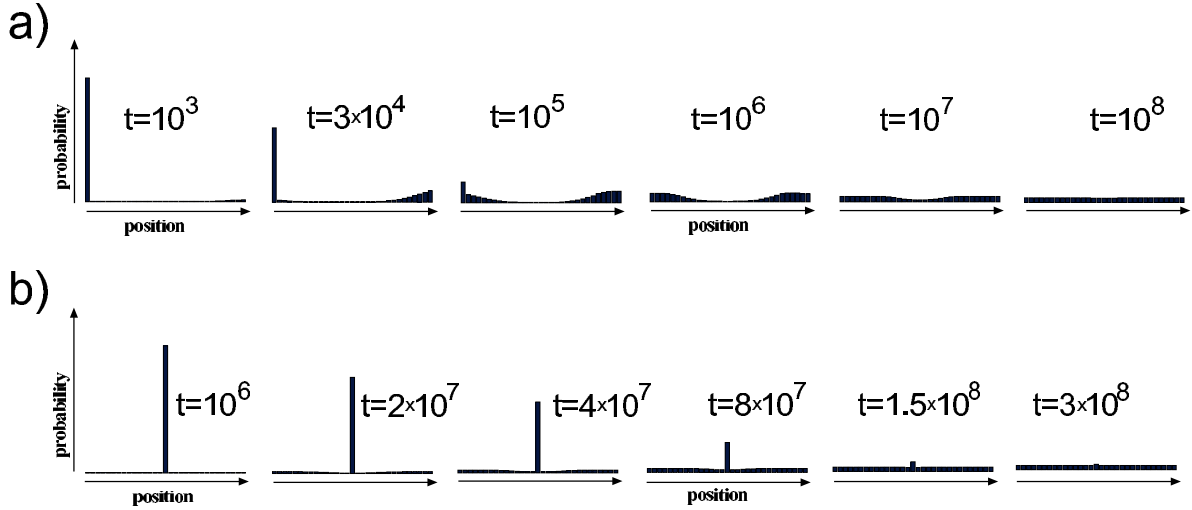}
\caption{Relaxation dynamics of two initial states of nucleosome
positions\ on a longer DNA segment (147 + 300 bp): a) end
positioned and b) centrally positioned initial species. Note the
initial difference in relaxation timescales for a) and b) (which
are due to different loop energies involved).}
\end{figure*}

\begin{figure*}
\includegraphics*[width=16cm]{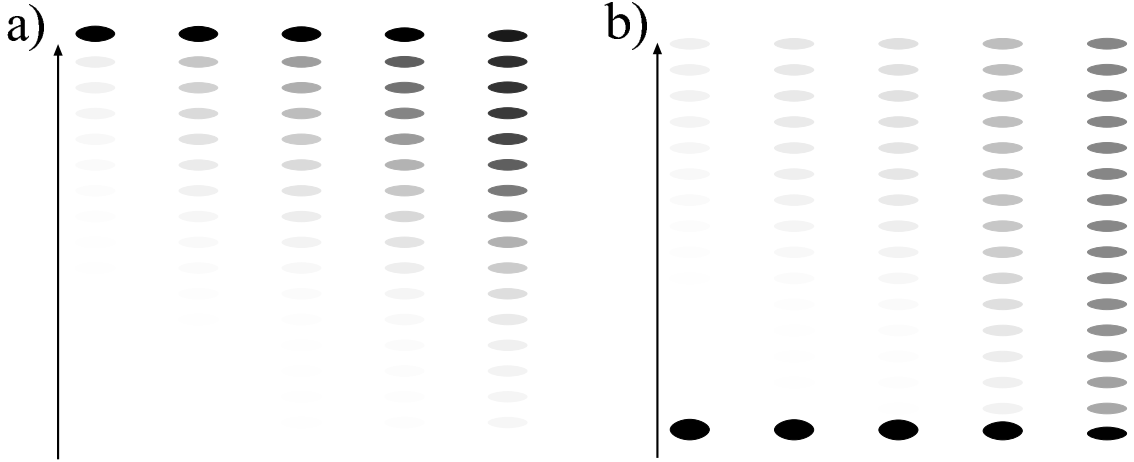}
\caption{The (1-D) gel electrophoresis signatures simulated for
the relaxation dynamics of the two initial species from Fig.9. a)
End positioned (lanes 1-5 corresponding to incubation times
(1,2,3,10,50)$\times 10^{4}$ $C_{A}^{-1}$ ) and b) centrally
positioned (incubation times (1,2,3,10,50)$\times 10^{6}$
$C_{A}^{-1}$).}
\end{figure*}

In the case of longer DNA (but still not entropic segments) like
the 147+300 bp segment in Figs. 9 and 10, similar effects as for
the short segments are expected but with significantly faster
relaxation times by typically 2-3 orders of magnitude as compared
to the corresponding short segment populations. The corresponding
(simulated) electrophoretic gels are shown in Fig.~10 where for
the centrally positioned case (Fig 10b) the ''population gap''
effect is even more pronounced than in the short segment case.

For even longer DNA\ segments we expect the gap effect to persist
(data not shown) and the optimal jump size to be around 2-3$\times
l_{P}$ corresponding to the free energy minimum in Fig.~5. For
very long DNA segments, the nucleosome repositioning behavior
implied by the big-loop-mechanism becomes strongly non-local which
contrasts a local diffusive motion as expected from cork-screwing
motion (cf. Refs. \cite{Beard,Spadafora,Pennings,Meersseman}) or
small loop repositioning as considered by Ref.~\cite{Schiessel}.
As mentioned above, this superdiffusive behavior has diverging
moments which implies strongly enhanced nucleosome transport along
very long DNA pieces. However such an ideal superdiffusion of
nucleosomes could hardly occur {\it in vivo} because free DNA
segments between subsequent nucleosomes (DNA linkers) are never
longer than $\sim O\left( l_{P}\right) $. Furthermore the
neighboring nucleosomes might be a significant barriers (if not
for loop formation then) for loop migration around the nucleosome
which is an indispensable event for loop-mediated repositioning.

\section{Conclusions and Discussion}

In this study we examined a possible mechanism for the
repositioning of nucleosomes along DNA which is based on the
formation and diffusion of intranucleosomal loops. The most
important outcome of this study is the prediction of two classes
of loops that might occur: (1) small 10bp-loops and (2) large
loops with a wide distribution of stored lengths with a weak peak
at roughly two times the DNA persistence length.

The small loops were already discussed in Ref.~\cite{Schiessel}
and led to the prediction of repositioning steps of 10bps.
Furthermore, the repositioning time should be of the order of an
hour, a consequence of the large activation energy required to
form a loop. This might explain the strong temperature dependence
of the typical repositioning time \cite{Meersseman}. In fact, by
lowering the temperature from $37^\circ$ to $4^\circ C$ no
redistribution within one hour was detected in that experiments.
Assuming a loop formation energy of $23k_{B}T$ one finds indeed a
slowing down of this process by factor of 13.

On the other hand, the large loop repositioning considered here
turns out to be energetically much more favorable. Loops with an
extra length of $2l_{P}$ have an energy that is roughly
$12$-$13k_{B}T$ smaller than that of a 10bp-loop. To a certain
extend this is because such loops can have a very small nucleosome
opening angle by forming crossed loops but the main contribution
stems from the significantly decreased DNA bending energy. One
therefore expects that repositioning via large loops should be the
dominant process on sufficiently large DNA pieces and that the
typical times are much shorter than the one for small loop
repositioning (say, of the order of minutes).

So far, however, the experiments did not report such events.
Meerseman et al. \cite{Pennings,Meersseman}, for instance, found
on short DNA pieces of 207bps length results that are consistent
with 10bp repositioning -- as we would expect for such short DNA
fragments. However, when they redid the experiment with a 414bp
long piece, a tandem repeat of the 207bp DNA, their analysis of
the complicated band patterns observed in 2D gel electrophoresis
did not show any indication that the nucleosome was able to move
from one half to the other.

Hence, the question arises if the repositioning observed in these
experiments was facilitated via the loop mechanism or if it
occurred via a different process. An analysis of the results is
made especially difficult by two complications: (a) the
nucleosomes seem to prefer to sit on the ends of the DNA fragments
and (b) most of the experiments use strong positioning sequences
(like the 5S rDNA sequence). This means that, independent of what
the repositioning process might be, the nucleosomes have certain
preferred positions and these might obscure the underlying
repositioning process.

With regard to this fact, let us consider two other repositioning
mechanisms
that one could imagine. The first one is that the nucleosome detaches {\it %
completely} from the DNA and attaches at some other position (or
even a different DNA molecule). This process, however, seems to be
excluded by two facts (among others). First that no repositioning
from one half to the other of the 414bp DNA or to competitor DNA
fragments was observed \cite{Pennings,Meersseman}. Secondly, once
completely detached from the DNA\ template the histone octamer
becomes unstable and disintegrates into a tetrameric and two
dimeric subunits which makes an effective nucleosome
reconstitution difficult.

The other mechanism could be a local screwing motion as already
suggested in Ref.~\cite{Pennings}. This process would lead to a
repositioning with one bp per step. The preponderance of 10bp
steps observed for the 5S rDNA experiments could then be explained
as being due to the fact that the positioning sequence prefers the
nucleosome rotationally positioned on one side of the DNA where it
can be easily bent around the octamer. Also 10bps (and even a few
multiples of 10bps) apart this effect can still be seen and hence
the nucleosome would prefer positions multiples of 10bps apart. To
our best knowledge, the experiments to date do not allow to
distinguish whether the 10bp repositioning works via small loops
or via cork-screwing.

It would be therefore important to perform experiments on DNA
pieces that do not provide the nucleosome with a preferred
rotational setting. In that case the 10bp footprint should
disappear if nucleosomes reposition themselves via cork-screwing.
It would also be important to perform experiments with rather long
DNA fragments since we expect that large-loop repositioning can be
detected in such systems.

Finally, we note that nucleosome repositioning {\it in vivo} is
facilitated via so-called chromatin remodeling complexes, huge
multi-protein complexes that harness energy by burning ATP
\cite{Remodeling Review 1,Remodeling Review2,Kornberg Review}.
There are basically two major classes: ISWI and SWI/SNF. The first
one seems to induce small scale repositioning which might work via
twisting DNA that leads to a corkscrew movement as discussed
above. It might, however, also be possible that this complex
induces small loops on the nucleosome as recent experiments on
nicked DNA suggest \cite{Längst}. The other class of remodeling
complexes seems to induce large loop structures as they have been
observed recently via electron spectroscopy \cite{Bazett-Jones et
al.}. Whatever the details of the functions of these remodeling
complexes might be, it is tempting to speculate that they catalyze
and direct processes which might even take place when they are not
present -- like small loop and large loop formation as well as
screwing. In this case the computed looping energy (cf. Fig.~5)
and repositioning rates might give a first hint about ATP
requirements and the dynamics of enzymatic repositioning.

Another interesting and very prominent system known to mediate
nucleosome repositioning via loop formation is unexpectedly the
ubiquitous RNA-Polymerase (RNA-P). It is found to be able to
transcribe DNA through nucleosomes without disrupting their
structure, yet moving them {\it upstream} the DNA template, i.e.,
in the opposite direction of transcription \cite {Felsenfeld RNA-P
Review}. To rationalize this seemingly paradoxical finding
Felsenfeld et al. introduced a DNA looping model \cite {Felsenfeld
RNA-P Review} which assumes that the RNA-P crosses the nucleosome
in a loop. This would indeed explain the backwards directionality
of repositioning. An interesting question in this context is how
our intranucleosomal loops considered above relate to those formed
by the RNA-P. Can we say something about the repositioning
distance distribution, does the looping energy (Fig.~5) apply
here? The geometry of RNA-P - DNA complex on a nucleosome is
certainly different from the simple loop case, as ingoing and
outgoing DNA from RNA-P enclose a (rather soft yet) preferential
angle of $\approx 100^{\circ }$ (dependent on RNA-P type, cf.
Refs. \cite{RNAP ANGLE 1,RNAP ANGLE 2,RNAP ANGLE 3}). The latter
facilitates the loop formation as the free DNA has to bend less to
fold back onto the octamer surface. Besides the apparent
differences from the ''naked'' intranucleosomal loops problem, a
slight generalization of our present model which incorporates the
preferential RNA-P ''opening'' angle can be performed within the
same mathematical framework developed here. It would be
interesting to compute the resulting nucleosome transfer distance
on short and long DNA templates in an analogous manner as
performed above. An outcome of such a study could be, for
instance, an answer to questions like:\ what is the highest linear
nucleosomal density in polynucleosomal arrays, up to which
nucleosomes are not to be removed from the DNA template (due to
loop formation and nucleosome transfer prohibited by the
neighboring nucleosome) during transcription.

Such fundamental biological questions make a further elaboration
of intra-nucleosomal loop theory, its generalization to different
loop geometries, and finally its application to different loop
creating proteins (SWI/SNF, RNA-P) an intriguing task for future
work.

\section{Appendix: The Circle-Line Approximation}

Although the Kirchhoff's analogy provides us with essentially
analytic solutions for the rod deformed in plane, the occurrence
of boundary conditions (like Eqs. \ref{Constraint1} and
\ref{Constraint2}) prevents us in most cases from obtaining
analytical expressions of all the parameters characterizing the
solution (like $\sigma $ and $m$ above). To overcome this problem,
we suggest here a simple geometric approximation scheme which will
prove to be useful in obtaining analytic results for loops within
a reasonable accuracy (usually with a deviation of 5-15\% from the
exact numeric results).

\begin{figure}
\includegraphics*[width=8cm]{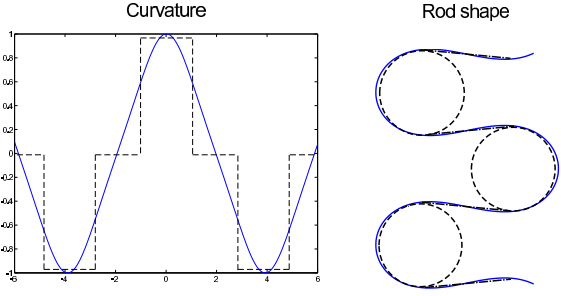}
\caption{The {\it circle-line approximation} for planar rods. The
curvature of an equilibrium rod shape (cn-function, cf.
Eq.~\ref{Curvature}) is approximated by a periodic sequence of
step-functions. The latter corresponds to an approximation of the
rod shape by a sequence of straight lines ($\kappa =0$) and
circles ($\kappa =$const.) glued together in a smooth manner
(continuous tangents).}
\end{figure}

The main idea is the following. The curvature and the energy (Eqs.
\ref{Curvature} and \ref{EbendExplicit}) of the loop contains the
cn$(\sigma |m)$ function which for $0<m<1$ has the typical
oscillatory behavior depicted in Fig.~11 (left). This suggests to
approximate the curvature function simply by a step function
consisting of an alternating sequence of negative, zero and
positive piecewise constant curvatures. Consequently the
corresponding rod shape (Fig.~11 right) is approximated by a
sequence of circles (positive / negative constant curvature) and
lines (zero curvature). An analogous approximation procedure can
also be performed in the case $m>1$ where the cn function has a
natural analytical continuation into a dn function with a modified
second argument (cf. Ref.~\cite{Abramowitz Stegun}).

Using this approximation ansatz several problems concerning planar
rods reduce to elementary geometry as seen from the following
simple but illustrative examples.

\begin{figure}
\includegraphics*[width=8cm]{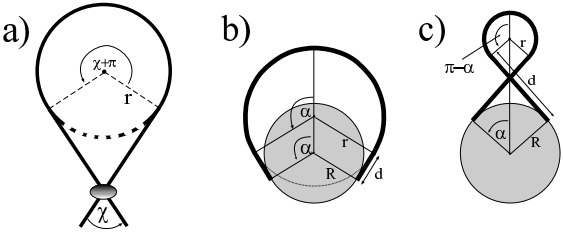}
\caption{Three applications of the circle-line approximation:
Problems with complex constraints reduce to simple geometries
leading to good approximations: a) the Yamakawa-Stockmayer angle
b) simple loops and c) crossed loops (see the Appendix text for
details).}
\end{figure}

\begin{figure}
\includegraphics*[width=8cm]{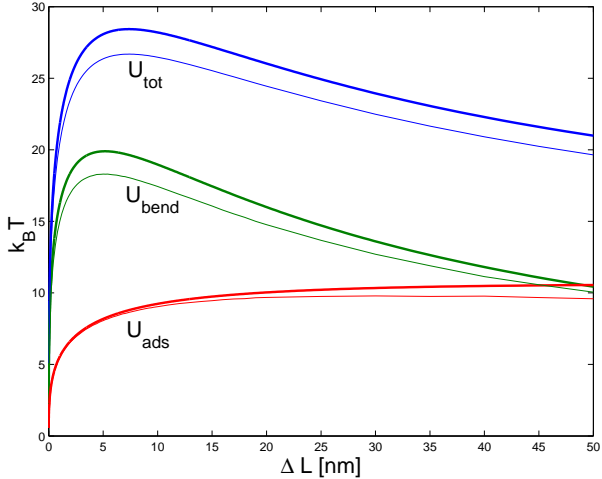}
\caption{Comparison of the adsorption and bending energy
contributions ($U_{ads} $ and $U_{bend}$) as well as the total
ground state energy $U_{tot}$ of the simple loop. The fat lines
represent the circle-line approximation (cf. Eq.~\ref{A1}) whereas
the thin lines show the corresponding exact expressions,
Eqs.~\ref{Etot} and \ref{EtotFinal} (thin line). The parameters
are $\varepsilon _{ads}=0.7 k_{B}T/nm$ and $A=50nm \times k_{B}T$
and $R=4nm$.}

\end{figure}

{\it 1) The Yamakawa-Stockmayer angle} \cite{Yamakawa Stockmayer}:
Two points on the rod are glued together without restricting the
orientation of the tangents, e.g., a protein connects two distant
points on DNA (cf. Fig.~12a). What is the preferred angle $\chi $
between the tangents in the ground state of the rod? By imposing a
fixed total rod length $L$ we have the simple constraint $L=\left(
2\cot \frac{\chi }{2}+\chi +\pi \right) r$ from which we can
eliminate $r$ and write the elastic energy of the configuration as
$U_{DNA}^{bend}=\frac{A}{L}(\chi +\pi )\left( 2\cot \frac{\chi
}{2}+\chi +\pi \right) $. Its minimization leads to the
transcendent condition $\chi _{\min }+\pi =\tan \chi _{\min }$
with the only relevant solution $\chi _{\min }\approx 77.5^{\circ
}$. The latter angle differs by $5\%$ from the exact result $\chi
_{\min }\approx 81.6^{\circ }$ (by Yamakawa and Stockmayer in
\cite{Yamakawa Stockmayer}) which is satisfactory regarding the
simplicity of the computation.

{\it 2) Simple and crossed loops} (Fig.~12 b,c){\bf :} We can
easily derive an approximate energy expression for simple /
crossed loops as a function of the excess length $\Delta L$ and
the opening angle $\alpha $. By applying simple geometry the
excess length constraint can be easily eliminated (the tangency
constraint is trivially fulfilled by the ansatz) and we arrive at
\begin{equation}
U_{simp}\left( \alpha \right) =2\alpha \left( A\frac{\tan \alpha -\alpha }{%
2R\left( \tan \alpha -\alpha \right) -\Delta L}+R\varepsilon
_{ads}\right) \label{A1}
\end{equation}
for simple loops and
\begin{eqnarray}
U_{cross}\left( \alpha \right) &=&2\alpha \left( A\frac{\pi +\tan
\alpha -\alpha }{\Delta L-2R\left( \tan \alpha -\alpha \right)
}+R\varepsilon
_{ads}\right)  \nonumber \\
&&+U_{def}\left( \alpha \right)  \label{A2}
\end{eqnarray}
for crossed loops where $A,R$ and $\varepsilon _{ads}$ defined as above and $%
U_{def}$ being the excluded volume interaction at the crossing
point, which is considered below (and applied in the main text as
Eq.~\ref{Udef}). We remark that the above expressions for
$U_{simp}$ and $U_{cross}$ are valid within certain $\alpha $
intervals which are given by the restriction $0<\alpha <\pi$ and
by the condition that the first terms in the brackets of Eqs.
\ref{A1} and \ref{A2} are positive (these are the necessarily
positive bending energy contributions in the two cases.)

These fairly simple expressions can now be used in the two cases
to obtain explicitly the ground state energies by minimizing Eq.
\ref{A1} and Eq.~\ref {A2} with respect to $\alpha $. For
instance, setting $U_{simp}^{\prime }\left( \alpha \right) =0$ we
obtain a transcendental equation for $\alpha $. We can now use the
fact that this condition is algebraic in $\Delta L$ so that we can
solve it for $\Delta L=\Delta L\left( \alpha \right)$. Thus
instead of finding $\alpha =\alpha(L)$ (which cannot be given in
an explicit form) we obtain explicitely its inverse:
\[
\frac{\Delta L\left( \alpha \right) }{R}=\frac{\left( 2-c\right)
G\left( \alpha \right) +cH\left( \alpha \right) }{1-c}
\]
\begin{equation}
+\chi \frac{\sqrt{\left[ \left( 2-c\right) G\left( \alpha \right)
+cH\left( \alpha \right) \right] ^{2}-4(1-c)G^{2}\left( \alpha
\right) }}{1-c} \label{A3}
\end{equation}
with the abbreviations
\begin{eqnarray*}
G\left( \alpha \right) &=&\tan \alpha -a \\
H\left( \alpha \right) &=&\alpha \tan ^{2}\left( \alpha \right)
\end{eqnarray*}
In Eq.~\ref{A3} the introduced dimensionless constant is $c=\left(
1+2R^{2}\varepsilon _{ads}/A\right) ^{-1}$ ($0<c<1$, and $c=0.69$ here) and $%
\chi $ is the sign accounting for different branches of the
$\alpha $ parametrized solution
\begin{equation}
\chi =\left\{
\begin{array}{c}
-1\text{ for }0\leq \alpha \leq \pi /2 \\
\pm 1\text{ for }\pi /2\leq \alpha \leq \alpha _{\max }\left(
c\right)
\end{array}
\right.
\end{equation}
Note that for $\alpha \leq \pi /2$ there is only one branch but
for $\alpha
>\pi /2$ we have two branches\footnote{%
The latter means that for $\pi /2\leq \alpha \leq \alpha _{\max }$
there are
two different excess loop lengths leading to the same (equilibrium)\ angle $%
\alpha $ ,i.e., with increasing $\Delta L$ the nucleosome angle
$\alpha $ opens but after passing some critical point on the
$\Delta L$ axis, it starts closing again.} ($\pm 1$) for $\Delta
L\left( \alpha \right) $. The maximal opening angle $\alpha _{\max
}\left( c\right) $ is obtained by setting the discriminant
(expression below the square root) in Eq.~\ref{A3} equal to $0$.

From Eq.~\ref{A3} together with Eq.~\ref{A1} we obtain an explicit
parametric representation of the minimal energy curve for simple
loops. A comparison of the approximate minimal energies
(Eq.~\ref{A3} and Eq.~\ref{A1}) with the exact minimal energy (cf.
also Fig.5 for $\Delta L\lesssim 60nm$) is shown in Fig.~13. We
find that the quantitative agreement is quite satisfactory taking
the simplicity of our ansatz into account. We note here that
analogous computations as we have shown for simple loops can be
performed for crossed loops as well.

For $\Delta L \rightarrow 0$ we find after an appropriate
expansion of $U_{simp}$ around $\alpha =0$ that the ground state
energy scales as $U_{simp}\sim \left( \Delta L/R\right) ^{1/3}$ in
agreement with Ref.~\cite {Schiessel}. Further we obtain the
excess length at which the loop ground state energy is maximal by
setting $\partial U_{simp}\left( \alpha \right) /\partial \alpha
|_{\alpha =\pi /2}=0$. From this follows the critical length
$\Delta L_{crit}$ as discussed in the main text (cf. Eq.
\ref{Lcrit}). This simple approximate expression for $\Delta
L_{crit}$ agrees within 2-15\% with the exact numerical result for
a wide range of adsorption energies with deviations becoming
larger for adsorption energies above $\varepsilon _{ads}=2.0$
$k_{B}T/nm$ (data not shown).
\begin{figure}
\includegraphics*[width=8cm]{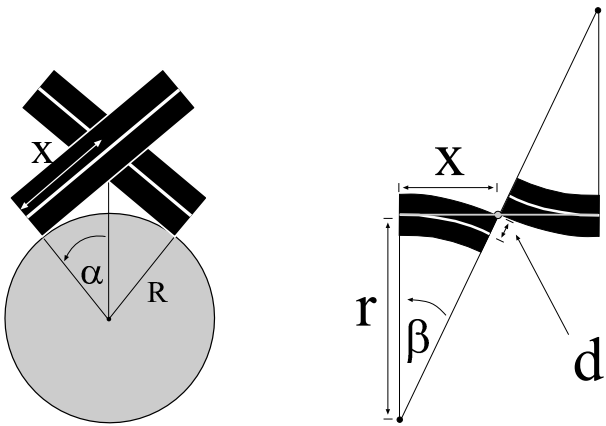}
\caption{The out of plane deflection of the incoming/outgoing DNA
due to excluded volume in the top projection (left) and seen from
the side (right). In the latter case (for the sake of visual
clarity) the two rods are depicted in a single plane, i.e.,
rotated around their contact point (grey dot).}
\end{figure}

{\it 3) The overcrossing potential for crossed loops }(Fig.
14){\bf :} The outgoing DNA\ path is perturbed out of the plane
due to the interaction with the ingoing DNA (and vice versa in a
symmetrical manner). Because of that our simple planar and phantom
model (no self interaction) needs modifications. Instead of
solving this (nonplanar) problem within the general theory of
self-interacting deformed rods as in Ref.~\cite{Coleman} (which is
a feasible but rather technical numerical task) we can treat the
out-of-plane deformation perturbationally. The first assumption we
make here is that the overall shape of the crossed loop does not
deviate much from a planar configuration though the orientation of
its (effective) plane might be slightly deflected from the
nucleosomal plane. Consequently the small perturbation out of the
plane and the deformation in plane essentially decuple into a sum
of two energy contributions as in Eq.\ref{A2}. Again by simple
geometry (cf. Fig 13), the second (out of the plane) term in
\ref{A2} can in first approximation be written as
\begin{equation}
U_{def}\left( \alpha \right) =\left\{
\begin{array}{c}
2A%
{\displaystyle{d\arctan \left( \frac{2dx\left( \alpha \right) }{x^{2}\left( \alpha \right) -d^{2}}\right)  \over x^{2}\left( \alpha \right) -d^{2}}}%
\text{ for }x\left( \alpha \right) >d \\
\infty \text{ otherwise}
\end{array}
\right.
\end{equation}
where $d\approx 1nm$ is the thickness of DNA and $x\left( \alpha
\right) :=R\tan \alpha $ the length of the crossed segment. In our
simple approximation the self-interaction energy diverges for
$x\longrightarrow d+0$ as $\frac{\pi }{2}A(x-d)^{-1}$ (extreme
deformation) and approaches zero for $x\longrightarrow \infty $ as
$4Ad^{2}x^{-3}$ (weak deformation).

We finally note that besides the above given examples it is
possible to apply the circle-line approximation to several other
standard problems of rod theory like the first and especially the
higher order Euler buckling instabilities to obtain qualitatively
the known results from buckling theory with very little effort.
Thus the circle-line approximation when applied appropriately
turns out to be very useful and generally allows computationally
inexpensive qualitative and quantitative insights into the
behavior of (planary) deformed rods.

\section{Acknowledgments}

We thank S. Mangenot, R. Bruinsma, W. M. Gelbart, J. Widom and R.
Everaers for useful discussions.

\end{document}